\pdfoutput=1
\documentclass[
    affil-it, 
    auth-lg, 
    british, 
    compression, 
    contents, 
    custom-packages={booktabs}, 
    references={bigone}, 
]{lib/preprint}

\let\oldblacksquare\blacksquare
\newcommand{\BBox}{{\textcolor{gray}{\mathop\oldblacksquare\nolimits}}} 
\newcommand{\BVbox}[1][]{%
    \ifthenelse{\equal{#1}{}}{%
        \mathrm{BV}^\BBox%
    }{%
        \mathrm{BV}^{\BBox,\,\mathrm{#1}}
    }%
}

\begin{document}
    
    \hypersetup{
        pdftitle = {Double-copying self-dual Yang–Mills theory to self-dual gravity on twistor space},
        pdfauthor = {Leron Borsten, Branislav Jurco, Hyungrok Kim, Tommaso Macrelli, Christian Saemann, Martin Wolf},
        pdfkeywords = {Double copy, self-dual Yang-Mills, self-dual gravity, color-kinematics duality, colour-kinematics duality, twistor space, holomorphic Chern-Simons theory},
    }
    
    \date{\today}
    
    \email{l.borsten@herts.ac.uk,branislav.jurco@gmail.com,hk55@hw.ac.uk,tmacrelli @phys.ethz.ch,c.saemann@hw.ac.uk,m.wolf@surrey.ac.uk}
    
    \preprint{EMPG--23--15,DMUS--MP--23/13}
    
    \title{Double-Copying Self-Dual Yang--Mills Theory\\[0.3cm] to Self-Dual Gravity on Twistor Space} 
    
    \author[a]{Leron~Borsten}
    \author[b]{Branislav~Jur{\v c}o}
    \author[c]{Hyungrok~Kim}
    \author[d]{Tommaso~Macrelli}
    \author[c]{Christian~Saemann}
    \author[e]{Martin~Wolf\kern0.1em}
    
    \affil[a]{Department of Physics, Astronomy, and Mathematics,\\University of Hertfordshire, Hatfield AL10 9AB, United Kingdom}
    \affil[b]{Mathematical Institute, Faculty of Mathematics and Physics,\\ Charles University, Prague 186 75, Czech Republic}
    \affil[c]{Maxwell Institute for Mathematical Sciences, Department of Mathematics,\\ Heriot--Watt University, Edinburgh EH14 4AS, United Kingdom}
    \affil[d]{Institute for Theoretical Physics, ETH Zurich, 8093 Zurich, Switzerland}
    \affil[e]{School of Mathematics and Physics,\\ University of Surrey, Guildford GU2 7XH, United Kingdom}
    
    \abstract{We construct a simple Lorentz-invariant action for maximally supersymmetric self-dual Yang--Mills theory that manifests colour--kinematics duality. We also show that this action double-copies to a known action for maximally supersymmetric self-dual gravity. Both actions live on twistor space and illustrate nicely the homotopy algebraic perspective on the double copy presented in~\cite{Borsten:2023ned}. This example is particularly interesting as the involved Hopf algebra controlling the momentum dependence is non-commutative and suggests a generalisation to gauged maximally supersymmetric self-dual gravity.}
    
    \acknowledgements{H.K.~and C.S.~were supported by the Leverhulme Research Project Grant RPG-2018-329. B.J.~was supported by the GA\v{C}R Grant EXPRO 19-28628X.}
    
    \datalicencemanagement{No additional research data beyond the data presented and cited in this work are needed to validate the research findings in this work. For the purpose of open access, the authors have applied a Creative Commons Attribution (CC BY) licence to any Author Accepted Manuscript version arising.}         
    \begin{body}
        
        \section{Introduction and results}

        It is a remarkable discovery of recent decades that a large class of natural gauge theories feature a hidden symmetry known as color--kinematics (CK) duality~\cite{Bern:2008qj,Bern:2010ue,Bern:2010yg}, which implies a surprising relation to gravitational theories, known as the double copy~\cite{Bern:2008qj,Bern:2010ue,Bern:2010yg} (for reviews, see~\cite{Carrasco:2015iwa,Borsten:2020bgv,Bern:2019prr,Adamo:2022dcm,Bern:2022wqg}). When confronted with such a fundamental new feature of quantum field theory, it is natural to examine the simplest non-trivial example of the phenomenon that nevertheless exhibits all interesting features. A strong contender for this title is (supersymmetric) self-dual Yang--Mills theory (SDYM; see e.g.~\cite{Popov:1998pc} for more details) --- which is even simpler than full $\caN=4$ supersymmetric Yang--Mills theory --- and its double copy, which is (supersymmetric) self-dual gravity (SDG; see e.g.~\cite{Krasnov:2016emc} for a review). 

        We present a remarkably simple twistor action for maximally supersymmetric SDYM theory that is reminiscent of the Leznov--Mukhtarov--Parkes action~\cite{Leznov:1986up,Leznov:1986mx,Parkes:1992rz} on space-time. Yet the action is Lorentz-invariant, manifests a kinematic Lie algebra and CK duality, and is based on a straightforward gauge-fixing from twistor space. It directly double-copies to another simple and known Lorentz-invariant twistor action for self-dual gravity~\cite{Mason:2007ct}.
        
        Our discussion of CK duality and the double copy makes use of the algebraic framework developed in~\cite{Borsten:2023ned,Borsten:2022aa} and based on ideas by~\cite{Reiterer:2019dys} (see also~\cite{Zeitlin:2009tj,Zeitlin:2014xma} as well as~\cite{Bonezzi:2022bse} for related work). Nevertheless, we have minimised the amount of mathematical prerequisites and omitted a review of the formalism; for a  detailed discussion of the general constructions, the reader should consult~\cite{Borsten:2023ned}. Below, we shall only expect some familiarity with the facts that a theory of fields taking values in a Lie algebra $\frg$ and with exclusively cubic interaction terms is encoded in a metric differential graded Lie algebra\footnote{`Graded' here always means $\IZ$-graded.} that factorises into a tensor product of $\frg$ and a metric differential graded commutative algebra. If the field theory comes with a kinematic Lie algebra, the latter can be promoted to a $\BVbox$-algebra by identifying a particular second-order differential operator $\sfb$ that defines the kinematic Lie bracket as a grade-shifted derived bracket. 
        
        We note that SDYM theory has been studied extensively in the context of CK duality and the double copy. In particular, the tree-level currents of SDYM theory in light-cone gauge were shown to exhibit CK duality in~\cite{Monteiro:2011pc} and to double copy to those of SDG. In the same paper, the kinematic Lie algebra of SDYM theory in light-cone gauge was identified with the area-preserving diffeomorphisms on $\IC^2$. More recently, we showed that holomorphic Chern--Simons theory on twistor space for (supersymmetric) SDYM theory manifests CK duality and the action implies CK duality for loop amplitudes in the maximally supersymmetric case~\cite{Borsten:2022vtg}. The corresponding full (un-gauge-fixed) kinematic Lie algebra is given by the Schouten--Nijenhuis-type Lie algebra of bosonic holomorphic multivector fields on twistor space, which reduces to the area-preserving diffeomorphisms on $\IC^2$, identified in~\cite{Monteiro:2011pc}, upon reducing to space-time and imposing light-cone gauge. Very recently~\cite{Bonezzi:2023pox}, a kinematic homotopy Lie algebra up to trilinear maps (encoding violations of the Jacobi identity up to homotopy) was derived directly from a gauge-invariant off-shell formulation of SDYM theory on space-time and put to a test in a double-copy construction of SDG in light-cone gauge. Again, by going to light-cone gauge, the kinematic Lie algebra of area-preserving diffeomorphisms on $\IC^2$ was also recovered. For further related work, including the classical double copy of self-dual solutions, see also~\cite{Bjerrum-Bohr:2012kaa,Monteiro:2013rya,Berman:2018hwd,White:2020sfn,Campiglia:2021srh,Krasnov:2021cva,Chacon:2021wbr,Monteiro:2022nqt,Ben-Shahar:2022ixa, Armstrong-Williams:2023ssz,Lipstein:2023pih,Easson:2023dbk,Farnsworth:2023mff}.        
        
        \section{Twistors, self-dual Yang--Mills theory, and self-dual gravity}
        
        \subsection{Self-dual Yang--Mills theory}
        
        \paragraph{Supersymmetric self-dual Yang--Mills theory.}
        Let $\frg$ be a metric Lie algebra with basis $\sfe_a$, structure constants $f_{ab}{}^c$, and metric $g_{ab}$. We set $f_{abc}\coloneqq g_{cd}f_{ab}{}^d$. The classical solutions to SDYM theory on Euclidean space $\IR^4$ are $\frg$-valued gauge potentials $A_\mu=A_\mu^a\sfe_a$ with self-dual field strength
        \begin{equation}\label{eq:SDYMeq}
            F_{\mu\nu}^a\ =\ \tfrac12\eps_{\mu\nu}{}^{\kappa\lambda}F^a_{\kappa\lambda}
            \ewith
            F_{\mu\nu}^a\ \coloneqq\ \partial_\mu A^a_\nu-\partial_\nu A^a_\mu+f_{bc}{}^aA^b_\mu A_\nu^c~,
        \end{equation}
        where we have coordinatised $\IR^4$ by $x^\mu$ with $\mu,\nu,\ldots=1,\ldots,4$, $\partial_\mu\coloneqq\parder{x^\mu}$, and $\eps_{\mu\nu\kappa\lambda}$ is the Levi-Civita symbol. An action for these configurations was given in~\cite{Chalmers:1996rq}, and there are supersymmetric extensions of both the equations of motion and the action from $\caN=1$ to $\caN=4$~\cite{Siegel:1992xp}.
        
        For the twistorial description of these solutions, it is convenient to switch to spinor notation. That is, we use the well known fact that the defining representation $\boldsymbol{4}$ of $\sfSpin(4)\cong\sfSU(2)_\rmL\times\sfSU(2)_\rmR$ decomposes as $\bm{4}\cong\bm{2}_\rmL\otimes\bm{2}_\rmR$, and we may set $x^{\alpha\dot\alpha}\coloneqq\sigma_\mu^{\alpha\dot\alpha}x^\mu$ where $\sigma_\mu^{\alpha\dot\alpha}$ are the sigma matrices with $\alpha,\beta,\ldots=1,2$ the chiral spinor indices and $\dot\alpha,\dot\beta,\ldots=\dot1,\dot2$ the anti-chiral ones. Then, the SDYM equation~\eqref{eq:SDYMeq} translates to
        \begin{equation}\label{eq:spinorSDYMeq}
            \eps^{\alpha\beta}(\partial_{\alpha\dot\alpha}A^a_{\beta\dot\beta}-\partial_{\beta\dot\beta}A^a_{\alpha\dot\alpha}+f_{bc}{}^aA^b_{\alpha\dot\alpha}A^c_{\beta\dot\beta})\ =\ 0~,
        \end{equation}
        where $\partial_{\alpha\dot\alpha}\coloneqq\parder{x^{\alpha\dot\alpha}}$ with $\eps_{\alpha\beta}=\eps_{\dot \alpha\dot\beta}=-\eps^{\alpha\beta}=-\eps^{\dot \alpha\dot \beta}$, $\eps_{\alpha\beta}=-\eps_{\beta\alpha}$, and $\eps_{12}=+1$, which implies $\eps_{\alpha\gamma}\eps^{\gamma\beta}=\delta_\alpha^\gamma$.
        
        CK duality of SDYM theory is most easily identified in the Leznov--Mukhtarov--Parkes form of the action~\cite{Leznov:1986up,Leznov:1986mx,Parkes:1992rz}, which is a result of adopting Leznov gauge, in which 
        \begin{equation}
            A_{\alpha\dot1}\ =\ \tfrac14\partial_{\alpha\dot2}\phi
            \eand
            A_{\alpha\dot2}\ =\ 0
        \end{equation}
        for $\phi$ some $\frg$-valued function on $\IR^4$ also called the prepotential. In this gauge,~\eqref{eq:spinorSDYMeq} reduces to
        \begin{equation}\label{eq:SDYMLeznov}
            \wave\phi^a+\tfrac12\eps^{\alpha\beta}f_{bc}{}^a (\partial_{\alpha\dot2}\phi^b)(\partial_{\beta\dot2}\phi^c)\ =\ 0
        \end{equation}
        with $\partial_{\alpha\dot\alpha}\partial^{\alpha\dot\alpha}=\frac12\partial_\mu\partial^\mu=\frac12\wave$. This equation follows variationally from the action
        \begin{equation}\label{eq:LMP_action}
            S^\text{LMP}\ \coloneqq\ \int\rmd^4x\left\{\tfrac12g_{ab}\phi^a\wave\phi^b+\tfrac{1}{3!}f_{abc}\eps^{\alpha\beta}\phi^c(\partial_{\alpha\dot2}\phi^a)(\partial_{\beta\dot2}\phi^b)\right\}.
        \end{equation}
        However, the equation of motion and the action are not Lorentz-covariant and not Lorentz-invariant, respectively.
        
        We can generalise this action to $\caN$-extended supersymmetric SDYM theory by extending $\IR^4$ to $\IR^{4|2\caN}$ by supplementing fermionic coordinates $\eta_i^{\dot\alpha}$ with $i,j,\ldots=1,\ldots,\caN$. The extended action reads as~\cite{Siegel:1992xp}
        \begin{equation}\label{eq:susyLMPAction}
            S^\text{LMP}\ \coloneqq\ \int\rmd^4x\,\rmd\eta^{\dot 2}_1\cdots\rmd\eta^{\dot 2}_\caN\left\{\tfrac12g_{ab}\phi^a\wave\phi^b+\tfrac{1}{3!}f_{abc}\eps^{\alpha\beta}\phi^c(\partial_{\alpha\dot2}\phi^a)(\partial_{\beta\dot2}\phi^b)\right\},
        \end{equation}
        in which $\phi$ is a superfield on $\IR^{4|2\caN}$ independent of $\eta_i^{\dot1}$. 
        
        Both actions exhibit CK duality of SDYM theory, as they feature a kinematic Lie algebra $\frK$ with Lie bracket
        \begin{equation}
            [\phi_1,\phi_2]_\frK\ \coloneqq\ \eps^{\alpha\beta}(\partial_{\alpha\dot2}\phi_1)(\partial_{\beta\dot2}\phi_2)~.
        \end{equation}
        
        \paragraph{Twistor basics.}
        As is well-known, $\caN$-extended supersymmetric SDYM theory has a twistorial reformulation in terms of holomorphic Chern--Simons theory~\cite{Ward:1977ta,Witten:2003nn,Popov:2004rb,Boels:2006ir}; see e.g.~\cite{Wolf:2010av} for a review. In the following, we summarise the underlying geometry.
        
        The twistor space $Z$ is the total space of the holomorphic vector bundle $\caO(1)\otimes\IC^{2|\caN}\rightarrow\IC P^1$. Geometrically, it parametrises all orthogonal almost-complex structures on $\IR^4$. We write $(z^A)=(z^\alpha,\eta_i)$ for the fibre coordinates and $\pi_{\dot\alpha}$ for the (homogeneous) base coordinates, where each of the indices $A,B,\ldots$ combines an $\alpha$ index and an $i$ index. Let us henceforth assume that $\caN$ is even; then $Z$ admits an anti-holomorphic involution $\tau:(z^A,\pi_{\dot\alpha})\mapsto(\hat z^A,\hat\pi_{\dot\alpha})$ with
        \begin{subequations}
            \begin{equation}
                \hat z^A\ \coloneqq\ \overline{z^B}C_B{}^A
                \eand
                \hat\pi_{\dot\alpha}\ \coloneqq\ C_{\dot\alpha}{}^{\dot\beta}\overline{\pi_{\dot\beta}}~,
            \end{equation}
            where
            \begin{equation}
                (C_A{}^B)\ \coloneqq\ \diag(C_\alpha{}^\beta,C_i{}^j)~,
                \quad
                (C_\alpha{}^\beta)\ \coloneqq\ \eps~,
                \quad
                (C_i{}^j)\ \coloneqq\ \unit_{\frac\caN2}\otimes\,\eps~,
                \eand
                (C_{\dot\alpha}{}^{\dot\beta})\ =\ -\eps~.
            \end{equation}
        \end{subequations}
        In the following, it will be useful to introduce the notation
        \begin{equation}
            |\pi|^2\ \coloneqq\ \eps^{\dot\alpha\dot\beta}\pi_{\dot\alpha}\hat\pi_{\dot\beta}\ =\ \pi_{\dot\alpha}\hat\pi^{\dot\alpha}~.
        \end{equation}
        The anti-holomorphic exterior derivative $\bar\partial$ on $Z$ can now be written as 
        \begin{subequations}
            \begin{equation}
                \bar\partial\ =\ \rmd\hat z^A\parder{\hat z^A}+\hat e^\pi\hat E_\pi~,
            \end{equation}
            where
            \begin{equation}
                \hat E_\pi\ \coloneqq\ |\pi|^2\pi_{\dot\alpha}\parder{\hat\pi_{\dot\alpha}}
                \eand
                \hat e^\pi\ \coloneqq\ \frac{\hat\pi^{\dot\alpha}\rmd\hat\pi_{\dot\alpha}}{|\pi|^4}~.
            \end{equation}
        \end{subequations}
        
        There is a diffeomorphism between $Z$ and $\IR^{4|2\caN}\times\IC P^1$. If we coordinatise the latter by $(x^{A\dot\alpha},\lambda_{\dot\alpha})=(x^{\alpha\dot\alpha},\eta^{\dot\alpha}_i,\lambda_{\dot\alpha})$ with $\lambda_{\dot\alpha}$ homogeneous coordinates on $\IC P^1$ and 
        \begin{equation}
            \hat x^{A\dot\alpha}\ \coloneqq\ \tau(x^{A\dot\alpha})\ =\ x^{A\dot\alpha}
            \quad\Leftrightarrow\quad
            x^{A\dot\alpha}\ =\ \overline{x^{B\dot\beta}}C_B{}^AC_{\dot\beta}{}^{\dot\alpha}~,
        \end{equation}
        then the diffeomorphism $Z\cong\IR^{4|2\caN}\times\IC P^1$ is given by
        \begin{equation}\label{eq:twistorDiffeo}
            \begin{gathered}
                (z^A,\pi_{\dot\alpha})\ =\ (x^{A\dot\alpha}\lambda_{\dot\alpha},\lambda_{\dot\alpha})
                \eand
                (x^{A\dot\alpha},\lambda_{\dot\alpha})\ =\ \left(\frac{z^A\hat\pi^{\dot\alpha}-\hat z^A\pi^{\dot\alpha}}{|\pi|^2},\pi_{\dot\alpha}\right).
            \end{gathered}
        \end{equation}
        Under this diffeomorphism, we obtain
        \begin{subequations}\label{eq:vectorFieldsAndForms}
            \begin{equation}
                \begin{aligned}
                    \left(\parder{\hat z^A},\hat E_\pi\right)\ &=\ \left(-\frac{1}{|\lambda|^2}\hat E_A,\hat E_\lambda+x^{A\dot\alpha}\lambda_{\dot\alpha}\hat E_A\right),
                    \\
                    (\rmd\hat z^A,\hat e^\pi)\ &=\ \left(-|\lambda|^2\hat e^A+|\lambda|^2x^{A\dot\alpha}\lambda_{\dot\alpha}\hat e^\lambda,\hat e^\lambda\right)
                \end{aligned}
            \end{equation}
            with
            \begin{equation}
                \begin{aligned}
                    (\hat E_A,\hat E_\lambda)\ &\coloneqq\ \left(\lambda^{\dot\alpha}\parder{x^{A\dot\alpha}},|\lambda|^2\lambda_{\dot\alpha}\parder{\hat\lambda_{\dot\alpha}}\right),
                    \\
                    (\hat e^A,\hat e^\lambda)\ &\coloneqq\ \left(-\frac{\hat\lambda_{\dot\alpha}\rmd x^{A\dot\alpha}}{|\lambda|^2},\frac{\hat\lambda^{\dot\alpha}\rmd\hat\lambda_{\dot\alpha}}{|\lambda|^4}\right).
                \end{aligned}
            \end{equation}
            We also set
            \begin{equation}
                E_A\ \coloneqq\ \parder{z^A}\ =\ \frac{1}{|\lambda|^2}\hat\lambda^{\dot\alpha}\parder{x^{A\dot\alpha}}~.
            \end{equation}
        \end{subequations}
        
        \paragraph{Holomorphic Chern--Simons theory.}
        Let $E\rightarrow Z$ be a complex vector bundle over $Z$ with vanishing first Chern class. Furthermore, let $\bar\nabla=\bar\partial+A$ be a $(0,1)$-connection on $E$ where $A$ is a $\frg$-valued $(0,1)$-form on $Z$. We also assume that there is a gauge in which $A$ has no anti-holomorphic fermionic directions $\rmd\hat\eta_i$ and depends holomorphically on the fermionic coordinates $\eta_i$; this is sometimes called Witten gauge~\cite{Witten:2003nn}. Then, the $\caN$-extended supersymmetric SDYM equation on $\IR^4$ is equivalent to the holomorphic Chern--Simons equation
        \begin{equation}\label{eq:hCS}
            \bar\partial A^a+f_{bc}{}^aA^b\wedge A^c\ =\ 0
        \end{equation}
        on $Z$~\cite{Ward:1977ta,Witten:2003nn,Popov:2004rb,Boels:2006ir}; see e.g.~\cite{Wolf:2010av} for a review. In the case of maximal ($\caN=4$) supersymmetry, $Z$ is a Calabi--Yau supermanifold, and~\eqref{eq:hCS} follows from varying a holomorphic Chern--Simons action on $Z$~\cite{Witten:2003nn}. See also~\cite{Sokatchev:1995nj} for a similar Chern--Simons-type action in harmonic superspace in this case.
        
        This holomorphic Chern--Simons formulation manifests a gauge-invariant, off-shell kinematic Lie algebra and CK duality directly at the level of the action and further extends CK duality to the loop level as explained in~\cite{Borsten:2022vtg}. For $\caN<4$, the holomorphic Chern--Simons form of the equation of motion still implies CK duality for the tree-level currents.
        
        \paragraph{Twistorial prepotential action.}
        Given that we have both the Siegel action~\cite{Siegel:1992xp} as well as the prepotential action~\eqref{eq:LMP_action} for SDYM theory on space-time, it is natural to ask if, besides the holomorphic Chern--Simons action, there is also a twistorial prepotential action.
        
        To derive such an action, we write the holomorphic Chern--Simons equation~\eqref{eq:hCS} for $A=\rmd\hat z^\alpha A_\alpha+\hat e^\pi A_\pi$ as
        \begin{equation}\label{eq:hCS2} 
            \begin{aligned}
                \parder{\hat z^\alpha}A^a_\beta-\parder{\hat z^\beta}A^a_\alpha+f_{bc}{}^a A^b_\alpha A^c_\beta\ &=\ 0~,
                \\
                \hat E_\pi A^a_\alpha-\parder{\hat z^\alpha}A^a_\pi+f_{bc}{}^a A^b_\pi A^c_\alpha\ &=\ 0~.
            \end{aligned}
        \end{equation}
        Since we have assumed that $E$ has vanishing first Chern class, we may work in the axial gauge
        \begin{subequations}
            \begin{equation}\label{eq:gaugehCS}
                A^a_\pi\ =\ 0~.
            \end{equation}
            In this gauge, the gauge potential has prepotentials $\phi^a$ and $\psi^a$, which are $\frg$-valued functions of weight~$2$ and~$0$ on $Z$, respectively. In particular,
            \begin{equation}\label{eq:prepotentials}
                A^a_\alpha\ =\ \frac{1}{4|\lambda|^2}(E_\alpha\phi^a+\hat E_\alpha\psi^a)
            \end{equation}
            with $E_\alpha$ and $\hat E_\alpha$ defined in~\eqref{eq:vectorFieldsAndForms}. This can be seen by regarding this equation as a vector-valued differential equation. The determinant of the differential operator $(E_\alpha~\hat E_\alpha)$ is $\hat E_\alpha E^\alpha=-\frac14\partial_\mu\partial^\mu=-\frac14\wave$. After restricting to functions that do not blow up at infinity, the kernel of $\wave$ consists of the constant functions, which are irrelevant in $A^a_\alpha$. Hence, the differential equation always has a solution.
            
            Let us further restrict to solutions to the holomorphic Chern--Simons equations~\eqref{eq:hCS2}. These are holomorphic in $\pi_{\dot\alpha}$, and we can therefore impose Lorenz gauge along the fibres,
            \begin{equation}\label{eq:WoodhouseSDYM}
                E^\alpha A^a_\alpha\ =\ 0~.
            \end{equation}
            This further restricts the prepotential to $\psi^a\in\ker(\wave)$, and hence we can put $\psi^a=0$. Moreover, the fact that $A^a_\alpha$ is holomorphic in $\pi_{\dot\alpha}$ allows us to demand that $\phi^a$ is holomorphic in $\pi_{\dot\alpha}$.
            
            Altogether, we see that the solutions to the holomorphic Chern--Simons equations are captured by a $\frg$-valued function of weight~2 on $Z$ that depends holomorphically on the fermionic coordinates $\eta_i$ as well as $\pi_{\dot \alpha}$. Substituting\footnote{Note that this expression is very reminiscent of the Woodhouse representative, cf.~\cite{Woodhouse:id}.}
            \begin{equation}
                A^a_\alpha\ =\ \frac{1}{4|\lambda|^2}E_\alpha\phi^a
            \end{equation}            
            into~\eqref{eq:hCS2}, we obtain the remaining equation of motion 
        \end{subequations}
        \begin{equation}\label{eq:twistor_SDYM_eom}
            \wave\phi^a+\tfrac12\eps^{\alpha\beta}f_{bc}{}^a(E_\alpha\phi^b)(E_\beta\phi^c)\ =\ 0~.
        \end{equation}
        For maximal ($\caN=4$) supersymmetry, this equation follows from the variation of the action
        \begin{subequations}\label{eq:twistorActionSDYM}
            \begin{equation}
                S^\text{SDYM}\ \coloneqq\ \int\operatorname{vol}_\text{SDYM}\left\{\tfrac12g_{ab}\phi^a\wave\phi^b+\tfrac{1}{3!}f_{abc}\eps^{\alpha\beta}\phi^c(E_\alpha\phi^a)(E_\beta\phi^b)\right\},
            \end{equation}
            where
            \begin{equation}
                \operatorname{vol}_\text{SDYM}\ \coloneqq\ \rmd^4x\,\frac{\lambda^{\dot\alpha}\rmd\lambda_{\dot\alpha}\,\hat\lambda^{\dot\alpha}\rmd\hat\lambda_{\dot\alpha}}{|\lambda|^4}\,\rmd\eta_1\cdots\rmd\eta_4~.
            \end{equation}
        \end{subequations}
        This twistor action resembles the space-time action~\eqref{eq:susyLMPAction}, but it is manifestly Lorentz-invariant. As mentioned in the introduction, this twistor action appears to be new; we have not found a description of a similar action in the literature, not even for harmonic superspace. It could have been found from the single copy of the corresponding self-dual supergravity action, which we describe in \cref{ssec:SDG}.
        
        \paragraph{Relation to space-time.}
        The superfield expansion of $\phi^a$ reads as
        \begin{subequations}
            \begin{equation}
                \phi^a\ =\ A^a+\eta_i\,\chi^{ia}+\tfrac12\eta_i\eta_j\,W^{ija}+\tfrac1{3!}\eps^{ijkl}\eta_i\eta_j\eta_k\,\tilde\chi^a_l+\eta_1\eta_2\eta_3\eta_4\,\tilde A^a~,
            \end{equation}
        \end{subequations}
        and after performing the Penrose--Ward transform for the gauge potential $A^a=\frac{1}{4|\lambda|^2}\rmd\hat z^\alpha E_\alpha\phi^a$, we recover the expected degrees of freedom on $\IR^4$ as displayed in \cref{tab:SDYM}.
        
        \begin{table}[ht]
            \vspace*{10pt}
            \begin{center}
                \begin{tabular}{rccccc}  
                    \toprule
                    Field & $A^a$ & $\chi^{ia}$ & $W^{ija}$ & $\tilde\chi^a_i$ & $\tilde A^a$
                    \\
                    \midrule
                    Helicity & 1 & ${\textstyle\frac12}$ & 1 & $-{\textstyle\frac12}$ & $-1$ 
                    \\
                    Multiplicity & 1 & 4 & 6 & 4 & 1 
                    \\
                    \bottomrule
                \end{tabular}
                \caption{Space-time SDYM fields and their helicities and multiplicities.\label{tab:SDYM}}
            \end{center}
        \end{table}
        
        To relate the twistor action~\eqref{eq:twistorActionSDYM} to space-time, one can Kaluza--Klein-expand the scalar field in terms of spherical harmonics on $\IC P^1$ and then integrate over the sphere. As is often the case for CK-dual actions, there are infinitely many auxiliary fields in this expansion that enforce the equations of motion. Holomorphy in $\pi_{\dot \alpha}$ amounts to the equation $(\hat E_\lambda+x^{\alpha\dot\alpha}\lambda_{\dot\alpha}\hat E_\alpha)\phi^a=0$, which relates different terms in the Kaluza--Klein expansion. The latter is of the form
        \begin{equation}
            \phi^a\ =\ \lambda^{\dot\alpha}\lambda^{\dot\beta}\phi^a_{\dot\alpha\dot\beta}+\lambda^{\dot\alpha}\lambda^{\dot \beta}\frac{\lambda^{\dot\gamma}\hat\lambda^{\dot\delta}}{|\lambda|^2}\phi^a_{\dot\alpha\dot\beta\dot\gamma\dot\delta}+\cdots~,
        \end{equation}
        and setting $\phi^a_{\dot1\dot1}=\phi^a_\text{space-time}$, $\phi^a_{\dot 1\dot 2}=\phi^a_{\dot 2\dot 2}=0$, e.g., we recover equation~\eqref{eq:SDYMLeznov} from~\eqref{eq:twistor_SDYM_eom}.
        
        \subsection{Self-dual gravity}\label{ssec:SDG}
        
        There is an analogous picture for self-dual gravity, which we describe in the following.
        
        \paragraph{Supersymmetric self-dual gravity.}
        Let $(M,g)$ be a four-dimensional oriented Riemannian manifold with metric $g$. The self-dual gravity equation is an equation on the curvature for the Levi-Civita connection. In particular, for vanishing cosmological constant,\footnote{In the case of non-zero cosmological constant, one may modify the self-duality condition~\cite{Lipstein:2023pih}.} the SDG equation on a local patch $U\cong \IR^4$ of $M$ reads as
        \begin{subequations}\label{eq:SDG}
            \begin{equation}
                R_{\mu\nu\kappa}{}^\lambda\ =\ \tfrac{\sqrt{\det(g)}}{2}\eps_{\mu\nu}{}^{\rho\sigma}R_{\rho\sigma\kappa}{}^\lambda~,
            \end{equation}
            where 
            \begin{equation}
                R_{\mu\nu\kappa}{}^\lambda\ \coloneqq\ \partial_\mu\Gamma_{\nu\kappa}{}^\lambda-\partial_\nu\Gamma_{\mu\kappa}{}^\lambda+\Gamma_{\mu\kappa}{}^\sigma\Gamma_{\nu\sigma}{}^\lambda-\Gamma_{\nu\kappa}{}^\sigma\Gamma_{\mu\sigma}{}^\lambda
            \end{equation}
            is the Riemann curvature tensor, and 
            \begin{equation}
                \Gamma_{\mu\nu}{}^\kappa\ \coloneqq\ \tfrac12g^{\kappa\lambda}(\partial_\mu g_{\nu\lambda}+\partial_\nu g_{\mu\lambda}-\partial_\kappa g_{\mu\nu})
            \end{equation}
        \end{subequations}
        are the Christoffel symbols. Suppose now that $M$ also admits a spin structure. Then, we can pick a vierbein $e^{\alpha\dot\alpha}=\sigma_a^{\alpha\dot\beta}e^a$ on $U$ such that
        \begin{equation}
            g\ =\ \tfrac12\eps_{\alpha\beta}\eps_{\dot\alpha\dot\beta}e^{\alpha\dot\alpha}\otimes e^{\beta\dot\beta}~.
        \end{equation}
        The Riemann curvature tensor decomposes as
        \begin{subequations}
            \begin{equation}
                \begin{gathered}
                    R_{\alpha\dot\alpha\beta\dot\beta\gamma\dot\gamma}{}^{\delta\dot\delta}\ =\ R_{\alpha\dot\alpha\beta\dot\beta\gamma}{}^{\delta}\delta_{\dot\gamma}{}^{\dot\delta}+R_{\alpha\dot\alpha\beta\dot\beta\dot\gamma}{}^{\dot\delta}\delta_{\gamma}{}^{\delta}~,
                    \\
                    R_{\alpha\dot\alpha\beta\dot\beta\gamma}{}^{\delta}\ =\ \eps_{\alpha\beta}R_{\dot\alpha\dot\beta\gamma}{}^{\delta}+\eps_{\dot\alpha\dot\beta}R_{\alpha\beta\gamma}{}^{\delta}~,
                    \quad
                    R_{\alpha\dot\alpha\beta\dot\beta\dot\gamma}{}^{\dot\delta}\ =\ \eps_{\alpha\beta}R_{\dot\alpha\dot\beta\dot\gamma}{}^{\dot\delta}+\eps_{\dot\alpha\dot\beta}R_{\alpha\beta\dot\gamma}{}^{\dot\delta}
                    \\
                    R_{\alpha\beta\gamma}{}^{\delta}\ =\ C_{\alpha\beta\gamma}{}^{\delta}+\Lambda\eps_{\gamma(\alpha}\delta_{\beta)}{}^{\delta}~,
                    \quad
                    R_{\dot\alpha\dot\beta\dot\gamma}{}^{\dot\delta}\ =\ C_{\dot\alpha\dot\beta\dot\gamma}{}^{\dot\delta}+\Lambda\eps_{\dot\gamma(\dot\alpha}\delta_{\dot\beta)}{}^{\dot\delta}
                \end{gathered}
            \end{equation}
            with
            \begin{equation}
                \begin{gathered}
                    R_{\alpha\beta\dot\gamma\dot\delta}\ =\ R_{\dot\gamma\dot\delta\alpha\beta}~,
                    \quad
                    R_{\alpha\beta\dot\gamma\dot\delta}\ =\ R_{(\alpha\beta)(\dot\gamma\dot\delta)}~,
                    \\
                    C_{\alpha\beta\gamma}{}^{\delta}\ =\ C_{(\alpha\beta\gamma)}{}^{\delta}~,
                    \quad
                    C_{\alpha\beta\gamma}{}^{\gamma}\ =\ 0~,
                    \quad
                    C_{\dot\alpha\dot\beta\dot\gamma}{}^{\dot\delta}\ =\ C_{(\dot\alpha\dot\beta\dot\gamma)}{}^{\dot\delta}~,
                    \quad
                    C_{\dot\alpha\dot\beta\dot\gamma}{}^{\dot\gamma}\ =\ 0~.
                \end{gathered}
            \end{equation}
        \end{subequations}
        The components $C_{\alpha\beta\gamma}{}^{\delta}$ and $C_{\dot\alpha\dot\beta\dot\gamma}{}^{\dot\delta}$ constitute the self-dual and anti-self-dual parts of the Weyl tensor, and $\Lambda$ is the cosmological constant. The Ricci tensor is
        \begin{equation}
            R_{\alpha\dot\alpha\beta\dot\beta}\ \coloneqq\ R_{\gamma\dot\gamma\alpha\dot\alpha\beta\dot\beta}{}^{\gamma\dot\gamma}\ =\ -2R_{\alpha\beta\dot\alpha\dot\beta}+3\Lambda\eps_{\alpha\beta}\eps_{\dot\alpha\dot\beta}~,
        \end{equation}
        and the curvature scalar is then given by
        \begin{equation}
            R\ \coloneqq\ 2R_{\alpha\dot\alpha}{}^{\alpha\dot\alpha}\ =\ 24\Lambda~.
        \end{equation}
        The SDG equation~\eqref{eq:SDG} is now equivalent to requiring
        \begin{subequations}\label{eq:SDG2}
            \begin{equation}
                \Big\{~R_{\alpha\dot\alpha\beta\dot\beta\dot\gamma}{}^{\dot\delta}\ =\ 0~\Big\}
                \quad\Leftrightarrow\quad
                \Big\{~R_{\alpha\beta\dot\gamma}{}^{\dot\delta}\ =\ 0~,
                \quad
                C_{\dot\alpha\dot\beta\dot\gamma}{}^{\dot\delta}\ =\ 0~,
                \eand
                \Lambda\ =\ 0~\Big\}\,,
            \end{equation}
            and hence,
            \begin{equation}
                R_{\alpha\dot\alpha\beta\dot\beta\gamma\dot\gamma}{}^{\delta\dot\delta}\ =\ \eps_{\dot\alpha\dot\beta}C_{\alpha\beta\gamma}{}^{\delta}\delta_{\dot\gamma}{}^{\dot\delta}~.
            \end{equation}
        \end{subequations}
        
        It was shown in~\cite{Mason:1989ye} that~\eqref{eq:SDG2} is equivalent to the existence of volume-preserving vector fields $E_a=\sigma_a^{\alpha\dot\alpha}E_{\alpha\dot\alpha}$ whose Lie brackets satisfy
        \begin{equation}\label{eq:SDG3}
            \big[E_{\alpha(\dot\alpha},E_{\beta\dot\beta)}\big]\ =\ 0~.
        \end{equation}
        By Frobenius' theorem, we may now choose local coordinates in which
        \begin{subequations}
            \begin{equation}
                E_{\alpha\dot2}\ =\ \partial_{\alpha\dot2}~,
            \end{equation}
            and we may further fix a gauge such that
            \begin{equation}
                E_{\alpha\dot1}\ =\ \partial_{\alpha\dot1}+\tfrac14\eps^{\beta\gamma}(\partial_{\alpha\dot2}\partial_{\beta\dot2}\phi)\partial_{\gamma\dot2}
            \end{equation}
        \end{subequations}
        for $\phi$ a real-valued function on $U$. Then, the SDG equation~\eqref{eq:SDG3} reduces to
        \begin{equation}\label{eq:SDGPlebanski}
            \wave\phi+\tfrac12\eps^{\alpha\beta}\eps^{\gamma\delta}(\partial_{\alpha\dot2}\partial_{\gamma\dot2}\phi)(\partial_{\beta\dot2}\partial_{\delta\dot2}\phi)\ =\ 0~.
        \end{equation}
        This is Pleba\'nski's second heavenly equation~\cite{Plebanski:1975wn}, and it resembles the SDYM equation~\eqref{eq:SDYMLeznov} in Leznov gauge. 
        
        The equation~\eqref{eq:SDGPlebanski} generalises to $\caN$-extended supersymmetric SDG~\cite{Siegel:1992wd},
        \begin{equation}\label{eq:susySDGPlebanski}
            \wave\phi+\tfrac12\eps^{\alpha\beta}(-1)^{|A|}\Pi^{AB}(\partial_{\alpha\dot2}\partial_{A\dot2}\phi)(\partial_{\beta\dot2}\partial_{B\dot2}\phi)\ =\ 0~,
        \end{equation}
        where $\phi$ now is a superfield on $\IR^{4|2\caN}$ independent of $\eta_i^{\dot1}$. In addition, $(\Pi^{AB})\coloneqq\diag(\eps^{\alpha\beta},\Pi^{ij})$ with the rank of $\Pi^{ij}$ depending on how much of the R-symmetry group $\sfSO(\caN,\IC)$ is gauged: in the ungauged case, $\Pi^{ij}=0$. The equation~\eqref{eq:susySDGPlebanski} is variational, and follows from the action~\cite{Siegel:1992wd}
        \begin{equation}\label{eq:susyPlebanskiAction}
            S^\text{SP}\ \coloneqq\ \int\rmd^4x\,\rmd\eta^{\dot 2}_1\cdots\rmd\eta^{\dot 2}_\caN\left\{\tfrac12\phi\wave\phi+\tfrac{1}{3!}\eps^{\alpha\beta}(-1)^{|A|}\Pi^{AB}(\partial_{\alpha\dot2}\partial_{A\dot2}\phi)(\partial_{\beta\dot2}\partial_{B\dot2}\phi)\right\}.
        \end{equation}
        Note that~\eqref{eq:SDG3} also generalises to $\caN$-extended supersymmetric SDG~\cite{Wolf:2007tx}, and so does then the above derivation of~\eqref{eq:susySDGPlebanski}.
        
        \paragraph{Twistor description.}
        Like $\caN$-extended supersymmetric SDYM theory, also $\caN$-extended supersymmetric SDG enjoys a twistorial reformulation via Penrose's non-linear graviton construction~\cite{Penrose:1976js,Ward:1980am,Manin:1988ds,Merkulov:1991kt,Merkulov:1992qa,Merkulov:1992,Merkulov:1992b,Wolf:2007tx,Mason:2007ct}. Below we follow the treatment in~\cite{Wolf:2007tx,Mason:2007ct}. 
        
        By studying finite complex structure deformations on the twistor space $Z$, it was shown in~\cite{Mason:2007ct} that the $\caN$-extended supersymmetric SDG equation can be reformulated on $Z$ as a holomorphic Chern--Simons equation with the (infinite-dimensional) gauge group given by the holomorphic Poisson transformations. Concretely, we introduce the holomorphic Poisson structure 
        \begin{equation}
            [f,g]\ \coloneqq\ (-1)^{|A|(|f|+1)}\Pi^{AB}\parder[f]{z^A}\parder[g]{z^B}
        \end{equation}
        on $Z$, where $\Pi^{AB}$ is the tensor that already appeared in~\eqref{eq:susySDGPlebanski}. The $\caN$-extended supersymmetric SDG equation is then equivalent to~\cite{Mason:2007ct}
        \begin{equation}\label{eq:SDGHCS}
            \bar\partial h+\tfrac12[h,h]\ =\ 0~,     
        \end{equation} 
        where $h$ is a $(0,1)$-form on $Z$ of weight~2. Just as the holomorphic gauge potential in the SDYM setting, also $h$ is assumed to have no anti-holomorphic fermionic directions $\rmd\hat\eta_i$ and to depend holomorphically on the fermionic coordinates $\eta_i$. Note that~\cite{Mason:2007ct} also discusses the more general case of non-vanishing cosmological constant. For maximal ($\caN=8$) supersymmetry, equation~\eqref{eq:SDGHCS} follows from the variation of a holomorphic Chern--Simons action~\cite{Mason:2007ct}. It should be noted that in this case, however, the twistor space $Z$ is not a Calabi--Yau supermanifold; nevertheless, the weights of $h$ cancel appropriately so as to render the action well defined.
        
        \paragraph{Twistorial prepotential action.}
        In order to connect~\eqref{eq:SDGHCS} to~\eqref{eq:susySDGPlebanski}, we follow the closely our discussion of SDYM theory and write the holomorphic Chern--Simons equation~\eqref{eq:SDGHCS} as
        \begin{equation}\label{eq:SDGHCS2} 
            \begin{aligned}
                \parder{\hat z^\alpha}h_\beta-\parder{\hat z^\beta}h_\alpha+[h_\alpha,h_\beta]\ &=\ 0~,
                \\
                \hat E_\pi h_\alpha-\parder{\hat z^\alpha}h_\pi+[h_\pi,h_\alpha]\ &=\ 0~.
            \end{aligned}
        \end{equation}
        Considering the case of vanishing cosmological constant, we may impose the gauge
        \begin{subequations}
            \begin{equation}\label{eq:gaugeSDGHCS}
                h_\pi\ =\ 0~,
            \end{equation}
            just as for SDYM theory. We obtain prepotentials that, for solutions to~\eqref{eq:SDGHCS2}, we can further constrain by imposing Lorenz gauge along the fibres, $E^\alpha h_\alpha=0$, so that we arrive at 
            \begin{equation}\label{eq:WoodhouseSDG}
                h_\alpha\ =\ \frac{1}{4|\lambda|^2}E_\alpha\phi
            \end{equation}
        \end{subequations}
        for $\phi$ now a function of weight~4 on $Z$ that depends holomorphically on the fermionic coordinates $\eta_i$ as well as on $\pi_{\dot\alpha}$. Hence,~\eqref{eq:SDGHCS2} reduces to  
        \begin{equation}
            \wave\phi+\tfrac12(-1)^{|A|}\Pi^{AB}\eps^{\alpha\beta}(E_AE_\alpha\phi)(E_BE_\beta\phi)\ =\ 0~,
        \end{equation}
        where we have again used~\eqref{eq:vectorFieldsAndForms} as well as $\hat E_\alpha E^\alpha=-\frac14\wave$. For maximal ($\caN=8$) supersymmetry, this equation arises from variation of the action~\cite{Mason:2007ct}
        \begin{subequations}\label{eq:twistorActionSDG}
            \begin{equation}
                S^\text{SDG}\ \coloneqq\ \int\operatorname{vol}_\text{SDG}\left\{\tfrac12\phi\wave\phi+\tfrac{1}{3!}(-1)^{|A|}\Pi^{AB}\eps^{\alpha\beta}(E_AE_\alpha\phi)(E_BE_\beta\phi)\right\},
            \end{equation}
            where now
            \begin{equation}\label{eq:def_vol_SDG}
                \operatorname{vol}_\text{SDG}\ \coloneqq\ \rmd^4x\,\frac{\lambda^{\dot\alpha}\rmd\lambda_{\dot\alpha}\,\hat\lambda^{\dot\alpha}\rmd\hat\lambda_{\dot\alpha}}{|\lambda|^4}\,\rmd\eta_1\cdots\rmd\eta_8~.
            \end{equation}
        \end{subequations}
        This twistor action resembles the space-time action~\eqref{eq:susyPlebanskiAction}; however, again, it should be noted that~\eqref{eq:twistorActionSDG} is manifestly Lorentz invariant. A similar action (and the corresponding equation of motion) exists also on harmonic superspace~\cite{Karnas:1997it}.
        
        \paragraph{Relation to space-time.} The superfield expansion of $\phi$ reads as
        \begin{subequations}
            \begin{equation}
                \phi\ =\ g+\eta_i\psi^i+\eta_{ij}A^{ij}+\eta_{ijk}\chi^{ijk}+\eta_{ijkl}W^{ijkl}+\eta^{ijk}\tilde\chi_{ijk}+\eta^{ij}\tilde A_{ij}+\eta^i\tilde\psi_i+\eta\tilde g~,
            \end{equation}
            where
            \begin{equation}
                \eta_{i_1\cdots i_k}\ \coloneqq\ \tfrac{1}{k!}\eta_{i_1}\cdots\eta_{i_k}
                \eand
                \eta^{i_1\cdots i_{8-k}}\ \coloneqq\ \tfrac{1}{k!}\eps^{i_1\cdots i_8}\eta_{i_{9-k}}\cdots\eta_{i_8}~.
            \end{equation}
        \end{subequations}
        The Penrose--Ward transform of the field $h$ with this expansion substituted in then yields the correspondence between the various components and the SDG fields on $\IR^4$ displayed in \cref{tab:SDG}. 
        
        \begin{table}[ht]
            \vspace*{10pt}
            \begin{center}
                \begin{tabular}{rccccccccc}
                    \toprule
                    Field & $g$ & $\psi^i$ & $A^{ij}$ & $\chi^{ijk}$ & $W^{ijkl}$ & $\tilde\chi_{ijk}$ & $\tilde A_{ij}$ & $\tilde\psi_i$ & $\tilde g$
                    \\\midrule
                    Helicity & 2 & ${\textstyle\frac32}$ & 1 & ${\textstyle\frac12}$ & 0 & $-{\textstyle\frac12}$ & $-1$ & $-{\textstyle\frac32}$ & $-2$
                    \\
                    Multiplicity & 1 & 8 & 28 & 56 & 70 & 56 & 28 & 8 & 1 
                    \\
                    \bottomrule
                \end{tabular}
                \caption{Space-time SDG fields and their helicities and multiplicities.\label{tab:SDG}}
            \end{center}
        \end{table}
        
        \section{Colour--kinematics duality of self-dual Yang--Mills theory}
        
        Let us now show that the action $S^\text{SDYM}$ defined in~\eqref{eq:twistorActionSDYM} features CK duality. Algebraically, this is achieved by constructing a $\BVbox$-algebra, following the prescription of~\cite{Borsten:2022vtg,Borsten:2023ned}.
        
        \paragraph{Differential graded Lie algebra.}
        Let $\scS(m)$ denote the space of smooth functions of weight $m$ on $Z$ that are holomorphic in the fermionic coordinates $\eta_i$ as well as the bosonic coordinates $\pi_{\dot\alpha}$ and are bounded on $Z$. Recall that cubic actions correspond to metric differential graded Lie algebras, see~e.g.~\cite{Jurco:2018sby,Jurco:2019bvp}. In the case of the action~\eqref{eq:twistorActionSDYM}, we have the differential graded Lie algebra~$\frL^\text{SDYM}\cong\frL^\text{SDYM}_1\oplus\frL_2^\text{SDYM}$ concentrated in degrees~$1$ and~$2$ with the underlying cochain complex
        \begin{equation}
            \sfCh(\frL^\text{SDYM})\ \coloneqq\ 
            \Big(\!\!
            \begin{tikzcd}
                *\ar[r] & \underbrace{\frg\otimes\scS(2)}_{\coloneqq\,\frL^\text{SDYM}_1}\ar[r,"\mu_1"] & \underbrace{\frg\otimes\scS(2)}_{\coloneqq\,\frL^\text{SDYM}_2}\ar[r] & *
            \end{tikzcd}
            \!\!\Big)
        \end{equation}
        and differential $\mu_1|_{\frL^\text{SDYM}_1}\coloneqq\id_\frg\otimes\wave[-1]$, where $[k]$ for $k\in\IZ$ denotes an isomorphism combined with a cochain degree shift by $-k$; it simply changes the  cochain degree of an element by $-k$. It comes equipped with an invariant inner product, whose components vanish except between degrees $1$ and $2$, for which
        \begin{equation}
            \inner{\phi}{\chi}\ \coloneqq\ \int\operatorname{vol}_\text{SDYM}\,g_{ab}\phi^a\chi^b
        \end{equation}
        for all $\phi\in\frL^\text{SDYM}_1$ and $\chi\in\frL_2^\text{SDYM}$. The interactions are encoded in the Lie bracket $\mu_2:\frL^\text{SDYM}\times\frL^\text{SDYM}\rightarrow\frL^\text{SDYM}$, which vanishes except between two elements of degree $1$, for which
        \begin{equation}
            \mu_2(\phi_1,\phi_2)\ \coloneqq\ \sfe_c\otimes f_{ab}{}^c\eps^{\alpha\beta}(E_\alpha\phi^a_1)(E_\beta\phi^b_2)
        \end{equation}
        for all $\phi_{1,2}\in\frL^\text{SDYM}_1$.
        
        \paragraph{Colour-stripping.}
        The differential graded Lie algebra $\frL^\text{SDYM}$ naturally factorises as 
        \begin{equation}\label{eq:colour-stripping}
            \frL^\text{SDYM}\ \cong\ \frg\otimes\frC^\text{SDYM}~,
        \end{equation}
        where $\frg$ is the gauge Lie algebra and $\frC^\text{SDYM}=(\frC^\text{SDYM},\sfd,\sfm_2)$ is a differential graded commutative algebra; see~\cite{Borsten:2021hua} for a generic description of this procedure, which is referred to as colour-stripping in the physics literature. The cochain complex underlying $\frC^\text{SDYM}$ is concentrated in degrees $1$ and $2$,
        \begin{subequations}
            \begin{equation}
                \sfCh(\frC^\text{SDYM})\ \coloneqq\
                \Big(\!\!
                \begin{tikzcd}
                    *\ar[r] & \underbrace{\scS(2)}_{\coloneqq\,\frC^\text{SDYM}_1}\ar[r,"\sfd"] & \underbrace{\scS(2)}_{\coloneqq\,\frC^\text{SDYM}_2}\ar[r] & *
                \end{tikzcd}
                \!\!\Big)
            \end{equation}
            with $\sfd|_{\frC^\text{SDYM}_1}\coloneqq\wave[-1]$, i.e.~the colour-stripped $\mu_1|_{\frL^\text{SDYM}_1}\coloneqq\id_\frg\otimes\wave[-1]$. Its associative graded-commutative product $\sfm_2$ and inner product $\inner{-}{-}$ are
            \begin{equation}\label{eq:vertex}
                \sfm_2(\phi_1,\phi_2)\ \coloneqq\ \eps^{\alpha\beta}(E_\alpha\phi_1)(E_\beta\phi_2)
                \eand
                \inner{\phi_1}{\chi}\ \coloneqq\ \int\operatorname{vol}_\text{SDYM}\,\phi_1\chi
            \end{equation}
        \end{subequations}
        for all $\phi_{1,2}\in\frC^\text{SDYM}_1$ and $\chi\in\frC^\text{SDYM}_2$. 
        
        From a physicist's perspective, the decomposition~\eqref{eq:colour-stripping} amounts to colour-stripping. The differential and the product in the differential graded commutative algebra $\frC^\text{SDYM}$ encode the kinematic contributions to the inverse propagator and the interaction vertex of the theory.
        
        \paragraph{$\BVbox$-algebra and colour--kinematics duality.} The above differential graded commutative algebra can now be enhanced to a $\BVbox$-algebra. Note that the propagator in this theory is $P=\sfid_\frg\otimes \frac{\sfb}{\wave}$, where $\sfb$ is given by the shift isomorphism 
        \begin{equation}\label{eq:b_as_shift_iso}
            \sfb \ : \ \frC^\text{SDYM}_2 \ \xrightarrow{~[1]~} \frC^\text{SDYM}_1
        \end{equation}
        (and $\sfb$ is necessarily trivial otherwise). The operator $\sfb$ satisfies
        \begin{equation}
            [\sfd,\sfb]\ =\ \sfd\sfb+\sfb\sfd\ =\ \wave \eand \sfb^2=0~,
        \end{equation}
        and it is a second-order differential operator in the sense of~\cite{Akman:1995tm}, cf.~\cite{koszul1985crochet,Borsten:2023ned}. Hence, $(\frC^\text{SDYM},\sfd,\sfm_2,\sfb)$ forms a $\BVbox$-algebra~\cite{Reiterer:2019dys}, see also~\cite{Akman:1995tm,Borsten:2023ned}. 
        
        The extension of $\frC^\text{SDYM}$ from a differential graded commutative algebra to a $\BVbox$-algebra implies the existence of a kinematic Lie algebra $\frK$ with Lie bracket given by 
        \begin{equation}
            \{\Phi_1,\Phi_2\}\ \coloneqq\ \sfb\sfm_2(\Phi_1,\Phi_2)-\sfm_2(\sfb\Phi_1,\Phi_2)-(-1)^{|\Phi_1|}\sfm_2(\Phi_1,\sfb\Phi_2)
        \end{equation}
        for all $\Phi_{1,2}\in\frC^\text{SDYM}$~\cite{Reiterer:2019dys,Borsten:2022vtg,Borsten:2023reb}. Explicitly, we have here
        \begin{equation}
            \begin{gathered}
                \{\phi_1,\phi_2\}\ =\ \sfb(\sfm_2(\phi_1,\phi_2))\ =\ \eps^{\alpha\beta}(E_\alpha\phi_1)(E_\beta\phi_2)\ \in\ \frC^\text{SDYM}_1~,
                \\
                \{\phi_1,\phi^+_2\}\ =\ \sfm_2(\phi_1,\sfb\phi^+_2)\ =\ \eps^{\alpha\beta}(E_\alpha\phi_1)(E_\beta\phi^+_2)\ =\ \{\phi^+_2,\phi_1\}\ \in\ \frC^\text{SDYM}_2
            \end{gathered}
        \end{equation}
        for all $\phi_{1,2}\in\frC^\text{SDYM}_1$ and $\phi_2^+\in\frC^\text{SDYM}_2$. 
        
        In the Feynman diagram expansion, we can now either work with propagator $\frac{\sfb}{\wave}$ and vertex $\sfm_2(-,-)$ or we choose to re-assign the operator $\sfb$ to the vertex, so that we are left with a propagator $\frac{1}{\wave}$ and the Lie bracket $\{-,-\}$ as vertex, as explained in~\cite{Borsten:2023reb}. The latter picture renders CK duality for both currents and amplitudes of SDYM theory manifest.
        
        Altogether, we conclude that the tree-level currents of SDYM theory with an arbitrary amount of supersymmetry exhibit CK duality.\footnote{The tree-level amplitudes therefore do so as well, but these are trivial.}
        
        \section{Double copy from self-dual Yang--Mills theory to self-dual gravity}
        
        We now further follow the formalism of~\cite{Borsten:2023ned} to construct a double copy of the $\caN=4$ supersymmetric SDYM twistor action~\eqref{eq:twistorActionSDYM}. The result will be the ungauged version of the $\caN=8$ supersymmetric SDG twistor action~\eqref{eq:twistorActionSDG}. In the following, we make this connection algebraically rigorous to provide an explicit and easy-to-follow example of the formalism developed in~\cite{Borsten:2023ned}. 
        
        \paragraph{Hopf algebra.} In the formalism of~\cite{Borsten:2023ned}, we use a Hopf algebra in order to control the momentum dependence of fields, which is crucial in the identification of the correct field content of the double copy theory. An interesting new feature of the example at hand is now that this Hopf algebra is non-commutative. To control the momentum dependence on twistor space $Z\cong\IR^{4|2\caN}\times\IC P^1$, we can use the usual bosonic momentum operators $\partial_\mu$ on $\IR^4$ as well as a generator of $\frsu(2)$ together with the quadratic Casimir operator of $\frsu(2)$ to characterise the spherical harmonics on $\IC P^1$. The smallest Hopf algebra $\frH_Z$ that contains these is the vector space of constant coefficient differential operators on $\IR^4$ tensored with the universal enveloping algebra of $\frsu(2)$. Contrary to the examples discussed before, e.g.~in~\cite{Reiterer:2019dys} or~\cite{Borsten:2023ned}, this Hopf algebra is non-commutative.\footnote{The paper~\cite{Reiterer:2019dys} discusses the possibility of using the universal enveloping algebra, but does not actually use it in the main example.}
        
        \paragraph{Free fields.}
        The construction of the double-copied differential graded Lie algebra described in~\cite{Borsten:2023ned} starts from the restricted tensor product of the colour-stripped differential graded commutative algebras 
        \begin{equation}
            \hat \frC\ \coloneqq \ \frC^\text{SDYM}\otimes^{\frH_Z}\frC^\text{SDYM}~,
        \end{equation}
        which is spanned by elements $\phi_1\otimes \phi_2\in \frC^\text{SDYM}\otimes\frC^\text{SDYM}$, where $\phi_{1,2}\in \frC^\text{SDYM}$ with 
        \begin{equation}\label{eq:hopfcondition}
        \chi\acton \phi_1=\chi\acton \phi_2~.
                \end{equation}
        Here,  $\chi \acton \phi$ denotes the action of Hopf algebra elements  $\chi\in \frH_Z$ on colour-stripped (anti)fields $\phi\in \frC^\text{SDYM}$, where $\frC^\text{SDYM}$ is understood to be a $\frH_Z$-module. The condition~\eqref{eq:hopfcondition} ensures that the double-copied fields all are fields on a single copy of space-time, instead of producing a double-field-theory-like situation. As a result, the underlying chain complex reads as 
        \begin{equation}
            \sfCh(\hat \frC)\ \coloneqq\ 
            \Big(\!\!
            \begin{tikzcd}
                *\ar[r] & \underbrace{\scS(4)}_{\coloneqq\,\hat\frC_2}\ar[r] & \underbrace{\scS(4)~\oplus~\scS(4)}_{\coloneqq\,\hat\frC_3}\ar[r] & \underbrace{\scS(4)}_{\coloneqq\,\hat\frC_4}\ar[r] & *
            \end{tikzcd}
            \!\!\Big)\,.
        \end{equation}
        This is clearly not the chain complex of a field theory: fields and anti-fields are usually elements of degree~1 and~2. To remedy the situation, we will have to degree-shift the complex and truncate it further to the kernel of the operator $\hat \sfb_-$, where
        \begin{equation}
			\hat \sfb_\pm\ \coloneqq\ \sfb\otimes \sfid\pm\sfid\otimes\,\sfb
		\end{equation}
        with the operator $\sfb$ defined in~\eqref{eq:b_as_shift_iso}. This is akin to the familiar level-matching condition of string theory. The result is the chain complex
        \begin{subequations}
            \begin{equation}\label{eq:diff_DC}
                \sfCh(\frL^\text{SDG})\ \coloneqq\ 
                \Big(\!\!
                \begin{tikzcd}
                    *\ar[r] & \underbrace{\scS(4)}_{\coloneqq\,\frL^\text{SDG}_1}\ar[r,"\hat \mu_1"] & \underbrace{\scS(4)}_{\coloneqq\,\frL^\text{SDG}_2}\ar[r] & *
                \end{tikzcd}
                \!\!\Big)\,,
            \end{equation}
            where
            \begin{equation}
                \hat \mu_1\ \coloneqq\ \sfd\otimes \sfid + \sfid \otimes\, \sfd\ =\ \wave[-1]~.
            \end{equation}
        \end{subequations}
        This complex has the expected double-copied fields in degree~1 and a further copy as anti-fields in degree~2.
        
        \paragraph{Interactions.} 
        The interaction terms in the equation of motion are given by the Lie bracket of the kinematic Lie algebra of $\hat \frC$, restricted to $\frL^\text{SDG}$. Explicitly, the kinematic Lie bracket is given by
        \begin{equation}\label{eq:Lie_bracket_DC}
            [\hat\phi_1[1],\hat\phi_2[1]]\ =\ \hat \sfb_+\hat \sfm_2(\hat\phi_1,\hat\phi_2)
            \ =\ \eps^{\alpha\beta}\eps^{\gamma\delta}(E_\alpha E_\gamma\hat\phi_1)(E_\beta E_\delta\hat\phi_2)~,
		\end{equation}
        for all $\hat \phi_{1,2}[1]\in \frL^\text{SDG}$, where $\hat \sfm_2\coloneqq \sfm_2\otimes \sfm_2$ is a graded commutative product on $\hat \frC$. This is evidently the expected vertex that is obtained by double-copying the colour-stripped vertex~\eqref{eq:vertex} of SDYM theory. 

        \paragraph{Differential graded Lie algebra.} The differential graded Lie algebra $\frL^\text{SDG}$ that describes the double copy field theory is now given by the chain complex~\eqref{eq:diff_DC}, with the given differential $\hat \mu_1$ together with the grade-shifted bracket~\eqref{eq:Lie_bracket_DC}, i.e.~the graded antisymmetric product 
        \begin{equation}
            \hat \mu_2(\hat\phi_1,\hat\phi_2)\ \coloneqq\ [\hat\phi_1[1],\hat\phi_2[1]][-1]~.
		\end{equation}
        Hence, we obtain the equations of motion of SDG with none of the R-symmetry gauged.
        
        \paragraph{Action.} 
        An ingredient mostly ignored e.g.~in~\cite{Reiterer:2019dys}\footnote{The paper~\cite{Reiterer:2019dys} mentions the need for the metric in the loop case but does not develop it further.} and~\cite{Bonezzi:2022bse}, is the metric on $\frL^\text{SDG}$. This additional datum is crucial for the discussion of an action principle and scattering amplitudes, as without it, only the CK-duality of currents is ensured.        
        
        According to the prescription of~\cite{Borsten:2023ned}, we define
        \begin{equation}
            \inner{\hat\phi_1[1]}{\hat\phi_2[1]}_{\frL^\text{SDG}}\ \coloneqq\ (-1)^{|\hat\phi_1|}\inner{\wave^{-1}(\sfd \otimes\sfid-\sfid\otimes \sfd)\hat\phi_1}{\hat\phi_2}_{\hat\frC}
		\end{equation}
        for all $\hat\phi_{1,2}[1]\in \frL^\text{SDG}$. This provides a cyclic structure on $\frL^\text{SDG}$, i.e.~a metric invariant under the differential $\hat\mu_1$ and compatible with the bracket $\hat \mu_2$ in the usual manner, cf.~the familiar Cartan--Killing form. The dressing by $\wave^{-1}(\sfd \otimes\sfid-\sfid\otimes \sfd)$ heuristically reflects the familiar fact that in the double copy the numerator of the propagator is doubled, while the denominator is not. It yields the correct length dimensions in the action principle and reproduces the expected results in well-known examples. For a more detailed motivation, see~\cite{Borsten:2023ned}.
        
        In the case at hand, $\sfd = \wave[-1]$, and the operators in this inner product cancel to the shift isomorphism
        \begin{equation}
			\wave^{-1}(\sfd\otimes\sfid-\sfid\otimes \sfd)\ =\ [-1]\otimes\sfid-\sfid\otimes[-1]
		\end{equation}
        with $[-1]$ the shift isomorphism $[-1]:\frC_1^\text{SDYM}\xrightarrow{~\cong~}\frC_2^\text{SDYM}$. For further details, compare also the discussion of biadjoint scalar field theory in~\cite{Borsten:2023ned}, for which propagator and metric take very much the same form as in this case. 
        
        As also discussed in~\cite{Borsten:2023ned}, the doubling of space-time yields an infinite volume factor from the integral over the doubled bosonic directions that needs to be removed. Recall from above that we restricted the fields to the bosonic diagonal of this doubled space-time, and that the infinite volume factor originates from constancy of the Lagrangian along the bosonic off-diagonal directions.
        
        Schematically, we have
        \begin{equation}
            \int\operatorname{vol}_\text{SDYM}\otimes\operatorname{vol}_\text{SDYM}~~~\xrightarrow{~~~}~~~\operatorname{vol}(\IR^4\times\IC P^1)\int\operatorname{vol}_\text{SDG}~.
		\end{equation}
        Altogether, we are left with a metric uniquely characterised by 
        \begin{equation}
            \inner{\hat\phi_1}{\hat\phi^+_2}_{\frL^\text{SDG}}\ =\ \int \text{vol}_\text{SDG}\,\hat\phi_1\hat\phi^+_2~,
		\end{equation}
        where $\text{vol}_\text{SDG}$ was defined in~\eqref{eq:def_vol_SDG}. Together with the differential~\eqref{eq:diff_DC} and the Lie bracket~\eqref{eq:Lie_bracket_DC}, we recover the metric differential graded Lie algebra whose corresponding action is the scalar SDG action on twistor space~\eqref{eq:twistorActionSDG}.
        
        \paragraph{Remarks.}
        We note that the gauged version of the self-dual gravity twistor action~\eqref{eq:twistorActionSDG} may also be obtained as the double copy of self-dual Yang--Mills theory with another theory whose action is
        \begin{equation}
            \tilde S^\text{SDYM}\ \coloneqq\ \int\operatorname{vol}_\text{SDYM}\left\{\tfrac12g_{ab}\phi^a\wave\phi^b+\tfrac{1}{3!}f_{abc}(-1)^{|A|}\Pi^{AB}\phi^c(E_A\phi^a)(E_B\phi^b)\right\}.
        \end{equation}
        Choosing appropriate $\Pi^{AB}$, we can obtain an action in which an arbitrary amount of R-symmetry is gauged.
        
        If we are content with CK duality and double copy at the level of currents, we can also consider the corresponding non-maximally supersymmetric theories. All our constructions bar that of the metric go through as described above. If one wishes to work with an action for these and related theories, one can achieve this by replacing twistor space with a fattened complex manifold, as in~\cite{Saemann:2004tt}, or with a weighted projective space, as in~\cite{Popov:2004nk}.
        
        
        
    \end{body}
    
\end{document}